\documentclass{aa}
\usepackage{graphicx}
\begin{document}
    \title{A population of extreme mid-to-near-infrared sources:
    obscured AGN and dusty starbursts} 

   \author{P.H. Johansson \inst{1,2} \and P. V\"ais\"anen \inst{3,4}
   \and M. Vaccari \inst{5}}

   \offprints{P.H. Johansson, \\
   \email{phjohans@ast.cam.ac.uk}}

   \institute{Institute of Astronomy,
   Madingley Road, Cambridge, CB3 0HA, UK \and Observatory, P. O. Box
   14, FIN-00014 University of Helsinki, Finland 
   \and European Southern Observatory, Casilla 19001, Santiago, Chile
   \and Departamento de Astronom\'ia, Universidad de
   Chile, Casilla 36-D, Santiago, Chile \and Astrophysics Group,
  Blackett Laboratory, Imperial College, Prince Consort Road, London SW7 2AZ,
   UK } 
   
\date{Received / Accepted }
\authorrunning{Johansson, V\"ais\"anen \& Vaccari}
\titlerunning{EMNOs: obscured AGN and dusty starbursts}

\abstract{We present a sample of mid-infrared detected sources from
  the European Large Area ISO Survey (ELAIS) regions 
  characterised by strong mid-IR radiation with faint near-IR and
  optical counterparts.  These extreme mid-to-near-IR
  objects (EMNOs) are defined here by a flux ratio of $f_{\nu}
  (15\mu{\rm m}) /  f_{\nu} (2.2\mu{\rm m}) > 25$.  
This population is not obvious in deeper small area ISO surveys,
though it produces more than 20\% of the observed cosmic IR background
radiation (CIRB) at 15$\mu$m above 1 mJy.
 Near-future large area deep mid-IR surveys with the Spitzer
  Space Telescope, however, are bound to 
  uncover large amounts of these objects, which we argue to most
  likely be obscured AGNs, based on SED shapes and X-ray data.  
Very strong dusty starbursts at $z>1$ may also have high
  mid-to-near-IR flux ratios, but using the MIR/NIR and 
  FIR/MIR ratios these may be separated.  Most of our EMNOs appear to
  be ULIRGs, half are also extremely red objects (ERO).  A curious
  case of a low redshift, 
  less luminous object with a very young stellar population is also
  found.   We predict that the  
  simple broad band selection method makes EMNOs a useful window into
  high-redshift obscured nuclear activity and its sought after
  relation to star-formation, in a similar way that EROs have 
  been used to define samples of high-redshift early type galaxies.

} 

\maketitle

\section{Introduction}

The broadband photometric study of galaxies has turned out to be an
efficient method of discovering new classes of galaxies.  For example,
the extension of optical photometry into the near-infrared region
revealed a population of sources that was not represented in optical
surveys (Elston, Rieke \& Rieke \cite{El88}).
These objects were called ``Extremely Red Objects''
(EROs) because of their very red optical-infrared colours $(R-K>5, \,
I-K>4)$, and have become a useful tool in constraining galaxy
formation and evolution models in the redshift range $z=1-3$. 
 
The Infrared Space Observatory (\textit{ISO}) opened up 
wavelengths for large scale galaxy surveys further in the infrared
(e.g.\ Franceschini et al. \cite{Fr01}).
However, only the shorthly available wealth of mid- and
far-infrared photometric data obtained with the Spitzer Space
Telescope (Werner et al. \cite{We04}) will truly uncover the 
obscured history of the universe at $z=1-3$ for systematic
and statistical studies. 

Luminous and ultraluminous IR-galaxies, (U)LIRGs ($L_{\rm IR} > 
10^{11} - 10^{12} {\rm L}_{\sun}$) are a key population producing
most of the stellar energy output in the Universe since the
recombination era (Elbaz \& Cesarsky \cite{El03}).
On the other hand, several  
shortcomings of the generally successful and appealing hierarchical
scenarios of galaxy formation may be due to the poor understanding
of the AGN/galaxy formation and starburst interplay (Granato et
al. \cite{Gr04}).  Indeed, extreme IR-galaxies always seem to have AGN
contributions in varying degrees (e.g. Farrah et al. \cite{Fa03};
Serjeant et al. \cite{Se03}).  

In this Paper we present a sample of faint ISO detected mid-IR
galaxies which we argue to belong to these crucial classes of objects
needed to be understood before galaxy evolution and structure
formation in the Universe can be fully modelled.  We show that a
selection made by high mid-IR to near-IR flux ratio will  
uncover specifically sources with obscured AGN
activity, and possibly extremely high starformation, with the AGN
component growing with the afore mentioned ratio.  Nearly all the
sources found in this study are above the ULIRG limit.
We discuss sub-classes of these bright mid-IR objects to aid the
classification and study of IR-galaxies undoubtedly soon to be
cataloged by Spitzer 
in plentitude.  Cosmology used throughout is: $\Omega_{m}=0.3$,
$\Omega_{\Lambda}=0.7$ and $H_{0}=70 \ \rm km s^{-1} Mpc^{-1}$.

\section{The EMNO criterion}
\label{definition}

We search for objects with a flux ratio $f_{\nu}  (15\mu{\rm m}) /
f_{\nu} (2.2\mu{\rm m}) \equiv f_{15}/f_{K} > 25$, and dub these {\em 
extreme mid-to-near-IR objects}, EMNOs. 
The limit was chosen because the very strongest starburst galaxies,
such as Arp220, are expected to have approximately this 
flux ratio (e.g.\ Charmandaris et al. \cite{Ch02}), while AGN related
phenomena produce steeper NIR-MIR SEDs (Haas et al. \cite{Ha03};
Prouton et al. \cite{Pr04}).  

\begin{figure*}
\resizebox{17cm}{!}{\includegraphics{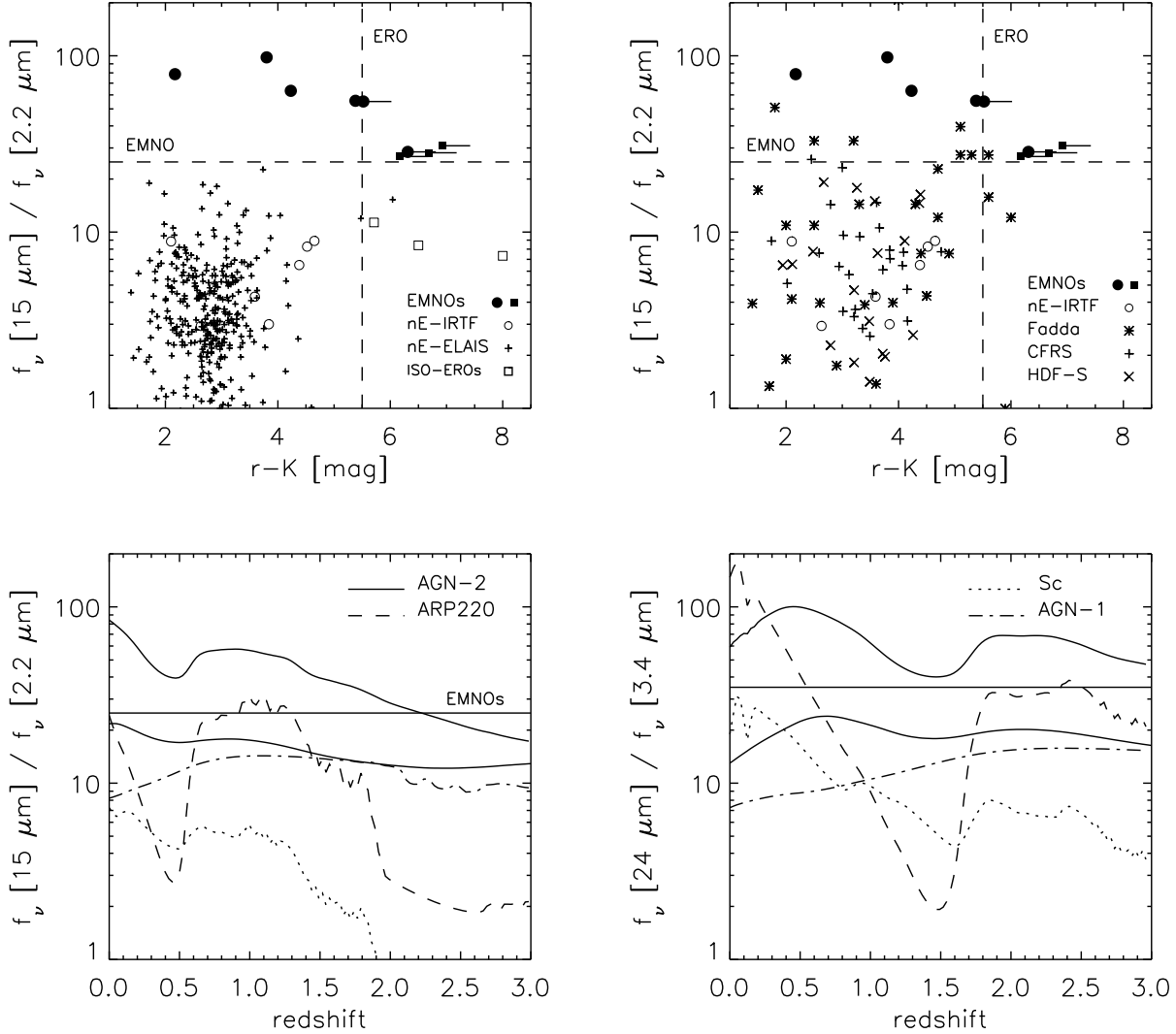}}
      \caption[]{Left panel: Upper panels: The $r-K$ colour vs. the 
15$\mu$m to $K$-band flux ratio.  Solid symbols show our
                 EMNOs, circles identifying IRTF EMNOs and squares
                 other ELAIS EMNOs. Non-EMNO ('nE') IRTF and ELAIS
                 sources are plotted as indicated, as well as data
                 from HDF-S (Mann et al. \cite{Ma02}), CFRS (Flores et
                 al. \cite{Fl99}), and X-ray selected MIR-sources from
                 HDF-N and Lockman hole (Fadda et al. \cite{Fa02}).
                 Lower panels:  model colours for
                 $f_{15}/f_{K}$ (ISO) and $f_{24}/f_{3.4}$ (Spitzer),
                 respectively, vs.\ redshift. Models are calculated 
                 using SEDs from GRASIL (Silva et 
                 al. \cite{Si98}) and Silva et al. (\cite{Si04}) 
                 (see text).
} 
         \label{iso_comb}
   \end{figure*}

The adopted criterion is tested in
Fig.~\ref{iso_comb} using ISO data and models.  In the upper panels we
plot the MIR/NIR flux ratio against the $r-K$ colour.  The small
crosses in the left panel are (extragalactic) sources from
European Large Area \textit{ISO} Survey (ELAIS; Oliver et
al. \cite{Ol00}) taken from the multiwavelength ELAIS Band-Merged
Catalogue\footnote{http://astro.imperial.ac.uk/Elais/Data\_release/}
(Rowan-Robinson et al. \cite{Row04}).  They are shown to mostly
consist of ``normal'' starforming galaxies (see also Pozzi et
al. \cite{Po03}; V\"ais\"anen et al. \cite{Va02}) and fall into the
lower left quadrant. The flux limit of ELAIS at $15\mu$m is
approximately 0.7 mJy, and $K\approx18$ and $r\approx24$ are the
limits in the other bands.

We also checked deeper ISO surveys, with sensitivity limits varying
between $\approx 0.1-0.4$ mJy, in the  
HDF-N, HDF-S, Lockman Hole, and CFRS fields (data from Aussel et
al. \cite{Au99}; Mann et al. \cite{Ma02}; Fadda et al. \cite{Fa02}; 
Flores et al. \cite{Fl99}). Data with appropriate bands available are
overplotted in the upper right panel of Fig.~\ref{iso_comb}.  Though
these deeper and small area surveys on average pick up slightly
redder sources than ELAIS, they do also clearly concentrate in the
same lower left region. The only population with $f_{15}/f_{K} > 25$ 
were found from the Fadda et al. (\cite{Fa02}; see also Franceschini
et al.\ \cite{Fr02}) X-ray selected Lockman data-set of ISOCAM
counterparts to XMM-Newton sources. {\em All} of the Fadda et
al. $f_{15}/f_{K} > 25$ sources are classified as obscured AGN
(typically with ${\rm N_{H}} > 10^{22} \ {\rm cm}^{-2}$) using X-ray
characteristics, though
6 out of 7 of them remained unclassified in the optical. 

Galaxy model colours, plotted in the lower panels of
Fig.~\ref{iso_comb}, are calculated using SEDs from the GRASIL model
(Silva et al. \cite{Si98}).  As seen, normal starforming galaxies
always have  
$f_{15}/f_{K}<10$.  IR-galaxies with very strong PAH emission such as
Arp220, have larger flux ratios, upto our EMNO criterion.  They are
rare in the local Universe. ULIRGs at higher redshift are expected to
be picked up more readily. However, K-correction (combined with strong
silicate absorption) hampers their detection at $z\sim0.5$ resulting
in an apparent gap in the MIR/NIR distribution in a flux-limited
sample.  

On the other hand, in sources where AGN activity heats the dust,
MIR/NIR ratio is typically large at any redshift.   
It is important to stress that the EMNO selection picks out
specifically obscured AGN.  Obscured nuclei with $N_H$ 
values in excess of $10^{22} \ {\rm cm}^{-2}$, are always EMNOs, often
with $f_{15}/f_{K}>100$ and the ratio increases with increasing $N_H$
(calculated from SEDs in Silva et al. \cite{Si04}), while e.g.\ Sy1
nuclei are never EMNOs (Fig.~\ref{iso_comb}, dash-dot curve).  We note
that in the parameterization of SEDs of Seyfert nuclei between 1 and
16$\mu$m  by 
Alonso-Herrero et al.\ (\cite{Al03}), our 
EMNO limit corresponds to their spectral index $\alpha_{IR}>1.7$,
where their Sy2s are found.   

Inclusion of the host galaxy properties complicates the models. 
In Fig.~\ref{iso_comb} the
higher plotted 'AGN-2' model is a 'custom scaled' composite of Arp220
SED and the nuclear SED of NGC1068 (latter taken from Silva et
al. \cite{Si04}), resulting in an EMNO. The lower solid curve is
calculated from composite type-2 QSO + host galaxy models of Silva et
al. (\cite{Si04}), using a combination giving a maximal
$f_{15}/f_{K}$: AGN X-ray luminosity $L_X > 10^{44} \ {\rm ergs \  
  s^{-1}}$ and $10^{23} < {\rm N_H} < 10^{24}$.  This implies that most
local AGN-hosting galaxies, where the SED templates were derived from,
are not quite extreme enough to be EMNOs.  The luminosity of the host
galaxy vs.\ the AGN plays an important role -- and thus can
potentially be studied -- with EMNO selection:  e.g.\ in the models of
Treister et al.\ (\cite{Tr04}), where an evolved elliptical is the
host of all obscured AGN, there are differences by factors of several
in $f_{15}/f_{K}$ compared to models used above with equivalent
choices of $L_X$ and $N_H$.  However, the general characteristics
of the models are similar, i.e.\ with host+AGN combination models
EMNOs are expected only from sources with strong AGN contribution,
$L_X > 
10^{43} \ {\rm  ergs \  s^{-1}}$, with the highest  $f_{15}/f_{K}$
ratios at $z>1$ produced by sources with  $N_H$ close to values
of $10^{23} \ {\rm cm}^{-2}$. 

When using the Spitzer MIPS/IRAC bands, Arp220 -like ultraluminous
starbursts have $f_{24}/f_{3.4}<35$ (except locally), and our
EMNO definition would then be $f_{24}/f_{3.4}>35$, corresponding to
Spitzer colour of $[3.6-24]>7.8$.  Indeed, a handful of such sources
are evident in early Spitzer results (e.g.\ Lonsdale et
al. \cite{Lo04}; Chary et al. \cite{Ch04}; Yan et al. \cite{Ya04}; Le
Floc'h et al. \cite{Lef04}) -- they are tentatively interpreted as
being luminous IR-galaxies where an AGN cannot be ruled out with the
available data.  The mentioned MIR/NIR flux ratio does not 
significantly change with the used IRAC band.  With $K$-band,
$f_{24}/f_{K}>60$ would be appropriate.  It is noteworthy that when
using the longer wavelength IRAC bands the 
expected AGN-2 flux ratio becomes increasingly similar to strong
starbursts, and thus $K$-band or $3.4\mu$m is best if the goal is 
to separate AGN from starformation.

\section{Observational data and the sample}

Our sample of near- and mid-IR detected galaxies presented here are
from the ELAIS fields N1 and N2. The $15\mu$m data, taken in the
ISOCAM LW3 band and reaching approximately 0.7 mJy, are from ELAIS
15$\mu$m Final analysis catalogue 
(Vaccari et al.\ \cite{Vac04}),  
available also within the multiwavelength ELAIS Band-Merged
Catalogue (Rowan-Robinson et al. \cite{Row04}).  The NIR observations,
however, are deeper than those available in the Band-Merged
Catalogue and are described in V\"ais\"anen \& Johansson (\cite{Va04b})
in detail -- essentially, we carried out a small-area 
survey to $K$~$<$~20 mag targeting faint ISOCAM sources
in approximately the range 1 to 3 mJy, without obvious counterparts in
DSS images.  
A total of 12 ISO sources were observed in this survey, and all were
found to have $K$-band counterparts. 
Observations were made with the 3.0-m NASA IRTF telescope
on Mauna Kea under photometric conditions (seeing varied between
0.7-1.0\arcsec) using the NFSCam $256 \times 256$ format InSb detector
array with a $0.3\arcsec$ pixel scale. Optical photometry in SDSS-like
filters is taken from the Isaac Newton Telescope Wide Field Survey 
(WFS)\footnote{http://www.ast.cam.ac.uk/$\sim$wfcsur/index.php} data
products (McMahon et al.\ \cite{Mc01}; Gonzales-Solares et al.\
\cite{Go04}).   

The MIR data is matched with the NIR-optical data-sets using a 
4\arcsec\ search radius around the ISOCAM detections. 
We do not expect spurious matches given that the probability of
finding a chance $K<19$~mag object within the search area is
$P\approx0.03$, and the ISOCAM detection error circle radius 
($1\sigma$) is better than 2\arcsec\ (Vaccari et al. \cite{Vac04}).

\begin{figure*}
\resizebox{18cm}{!}{\includegraphics{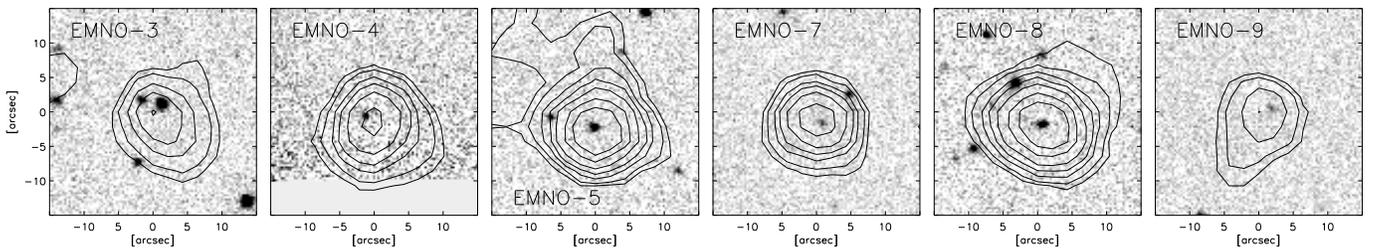}}
      \caption[]{The 15 $\mu$m contours overlaid on IRTF $K$-band
        images.  Contours designate 2,3,5,7,10,15, and 20$\sigma$
        SNR levels.}  
         \label{emno_maps}
   \end{figure*}

We find 6 EMNOs from the IRTF data, from the total of 12 matches between
ISO and IRTF data.  One of them is the fainter NIR counterpart of an
ISO object with two possible counterparts.  In addition, we
searched for EMNOs in the full published ELAIS catalogue (the ISO data
characteristics are the same as in our EMNO survey), which covers
a much wider area but utilises a significantly shallower NIR coverage:
three additional sources are found (EMNOs 1, 2, and 6).  The
photometry of all the 9 sources is presented in Table~\ref{table-phot},
and Fig.~\ref{emno_maps} shows the 6 EMNOs from our own $K$-band images
with 15$\mu$m contours overlaid.  

\begin{table*}
\scriptsize
\begin{center}
\begin{tabular}{crrrrcccccrcl}
            \hline
     \hline
     \noalign{\smallskip}
  Object   & $g'$ & $r'$ & $i'$ & $Z$ & $K$ &  $f_{15}$ & S/R &
 $\frac{MIR}{NIR}$ &  $z$ & SFR & $L_{IR}$ & SED  \\  
  (1)     & (2) & (3) & (4) & (5) & (6)  & (7) & (8) &  (9)  & (10)
     & (11) & (12) & (13)   \\   

            \noalign{\smallskip}
            \hline
            \noalign{\smallskip}

EMNO-1 ~ ELAISC15\_J160913.2+542320  &  $>25.00$ &  $>24.10$ &
$>23.20$ & $>22.00$ & 17.18  &    2.61 &  12.38  & 30.9 & (1.4) & 1090
& 12.7 & HR10 \\ 

EMNO-2 ~ ELAISC15\_J161024.4+542328  &  $>25.00$ &  $>24.10$ &
$>23.20$ & $>22.00$ & 17.42  &    1.91 &  12.19  & 28.2 & (1.4) & 870
& 12.6 & HR10 \\ 

EMNO-3 ~ ELAISC15\_J163531.1+410025  &  $>25.00$ &  $>24.10$ &  21.98 &
20.69 & 17.79 &    1.40 &  9.89  & 28.5  & $1.2^{+0.2}_{-0.4}$ & 170 & 12.0 & M82 \\

EMNO-4 ~ ELAISC15\_J163543.1+410750  &  $>25.00$ &  $>24.10$ &
$>23.20$ &  $>22.00$ & 18.58 &    1.66 &  16.86  & 55.1  & (1.5) & 300
& 12.1 & M82 \\ 

EMNO-5 ~ ELAISC15\_J163615.7+404759  &  23.90 &  22.85 &  21.32 &
21.17 & 17.47 &    2.78 &  31.24  &  55.6  & $0.9^{+0.1}_{-0.2}$  & 1000 & 12.7 &
Arp220 \\ 

EMNO-6 ~ ELAISC15\_J163748.1+412100  &  $>25.00$ &  $>24.10$ &
$>23.20$ & $>22.00$ & 17.93  &    1.14 &  6.65  & 26.9 &  (1.4)  & 550
& 12.4 & HR10 \\ 

\noalign{\smallskip}
\hline
\noalign{\smallskip}

EMNO-7 ~ ELAISC15\_J163515.6+405608  &  23.20 &  22.82 &  22.52 &
$>22.00$ & 19.02  &    2.14 &  19.88  &  98.0 &  $2.5^{+0.2}_{-0.8}$ &   &  &Sc 4Gyr\\

EMNO-8 ~ ELAISC15\_J163541.7+405900  &  20.66 &  19.95 &  19.62 &
19.99 & 17.78 &    3.69 &  28.32  &  78.6  & 0.188 &   &  &Sc 2Gyr\\

EMNO-9 ~ ELAISC15\_J163655.8+405909   & 22.92 &  22.96 &  23.02 &
$>22.00$ &  18.73 & 1.02  &  7.17  & 63.3 & 2.610 &   &  &Sc 3Gyr\\

\noalign{\smallskip}
\hline
\end{tabular}
\end{center}
\small
\caption[]{Photometry of the EMNOs.  Columns (2)-(5) give optical
  magnitudes (Vega) in 2.4\arcsec\ diameter apertures.  Our
  $K$-band (Vega) data (6) is determined in matched apertures, except 
  sources 1, 2, and 6 which use total $K$-band mags from the 
  ELAIS Band Merged
  Catalogue (Rowan-Robinson et al. \cite{Row04}).  Typical photometric
  errors are $\approx0.10$ mag for both optical and NIR data, and
  typical seeing was 1\arcsec\ for all data and no seeing corrections
  were applied to the magnitudes.
  (7)-(8) $15\mu$m flux and the corresponding 
  signal-to-noise ratio. The $f_{15}/f_{K}$ flux-ratio
  (9) is calculated from total fluxes.  (10) is the redshift: three
  decimals signal a spectroscopic redshift, one decimal a photometric
  determination (optical-NIR data) with HYPERZ $1\sigma$ errors, and a value 
  in parenthesis is a best-guess based on the over-all SED with typical
  errors of $z=\pm0.5$. 
  Columns (11)-(12) tabulate
  the SFR and total IR luminosity assuming the (10) redshift and SED
  given in (13).  The optical-NIR counterpart of EMNO-3 is one out of
  two possible sources (see Fig.~\ref{emno_maps}), 
the brighter one is a non-EMNO QSO at $z=1.15$.  
} 
\label{table-phot}
\end{table*}

\section{Discussion}

\subsection{Colours and colour-colour diagrams}

The MIR/NIR flux ratio vs. optical to NIR colour of our EMNOs are
shown as solid symbols in the upper panels of Fig.~\ref{iso_comb}.  It
is obvious that they are well separated from the bulk of MIR
galaxies.  Our non-EMNO IRTF/ISOCAM sources are plotted with open
circles.  Six of the EMNOs are also EROs, or very close to being EROs
(defined as $r-K>5.5$, see V\"ais\"anen \& Johansson \cite{Va04a}). As
is well known, EROs are thought to be either passively evolving
ellipticals or highly reddened dusty starbursts.  For comparison, we
overplot (open squares) three spectroscopically confirmed
ISO-detected EROs (Pierre et al. \cite{Pi01}; Smith et
al. \cite{Sm01}; Elbaz et al. \cite{El02b}).  None of them are EMNOs,
although the overall SED of especially HR10 (the reddest $r-K$ of the 
plotted three) is very similar to e.g.\ our EMNO-5 (see below). 

We note that all our EMNOs are part of the high $f_{15}/f_{r'}>10^2$ 
ratio population found by Gonzalez-Solares et al. (\cite{Go04}) and La
Franca et al. (\cite{La04}), mostly consisting of 15 $\mu$m sources
not identified in optical images.  Those EMNOs which are also EROs
have much more extreme MIR/optical ratios, $f_{15}/f_{r'}>10^3$.

\subsection{SEDs and redshifts}

\begin{figure*}
\resizebox{18cm}{!}{\includegraphics{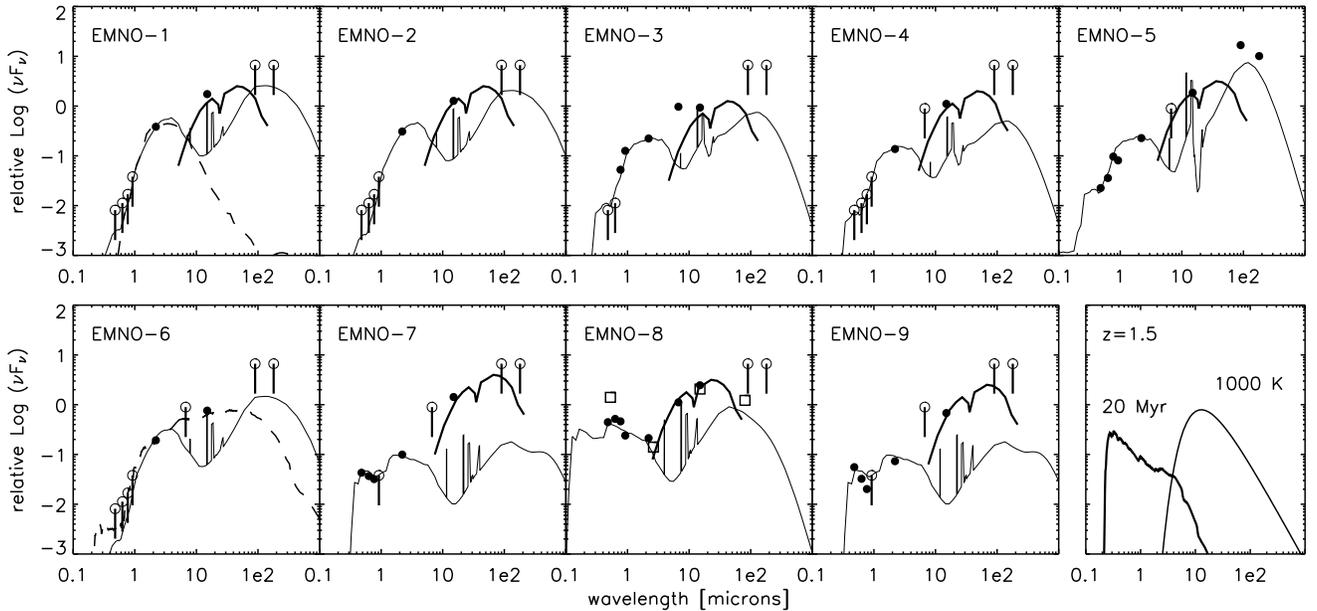}}
      \caption[]{Optical-to-FIR SEDs of our EMNOs. Solid circles show
      detections in $g', r', i', Z, K$ 
      and $6.7, 15, 90, 175 \mu$ bands, and the open circles the
      corresponding upper limits.  GRASIL models (thin line, Silva et 
      al. \cite{Si98}) given in Table~\ref{table-phot} are
      overplotted, as well as the nuclear SED 
      of NGC1068 (thick line, Silva et al. \cite{Si04}).  EMNO-1 is
      overplotted also with a 3Gyr elliptical at $z=2.5$ (dashed), and
      EMNO-6 with an obscured AGN model from Treister et
      al. (\cite{Tr04}, see text). The
      observed SED of SBS0335-052 is shown as squares with EMNO-8.  For
      reference, the lower right panel shows a 20 Myr 
      SED from Bruzual \& Charlot (\cite{Br93}) and a 1000 K
      blackbody. }    
         \label{emno_seds}
   \end{figure*}

Figure~\ref{emno_seds} shows the observed SEDs of our EMNOs.  
It is even more obvious than from the $r-K$ colour alone that there
are two kinds of EMNOs, those with optically red colours (EMNOs 1--6)
and those that are bluer (7--9). We discuss them separately.  

We fitted representative
IR-bright GRASIL models (Silva et al. \cite{Si98}), namely M82,
Arp220, and the extremely obscured starburst ERO 
HR10, to the optical--FIR SEDs of red EMNOs. Later ISO data has shown
the HR10 IR-SED to closely resemble that of the starburst ULIRG Arp220
(Elbaz et 
al. \cite{El02b}) but we chose here to retain the older GRASIL-SED to
represent the intermediate IR-luminosities between M82 and Arp220.   

We had spectroscopic redshifts for only two objects in our EMNO sample
and for the other sources we had to resort to photometric redshift methods. 
For the three EMNOs with optical detections we calculated photometric   
redshifts using the HYPERZ software (Bolzonella et al. \cite{Bo00}). For 
EMNO-3 we found a best fit of $z=1.2^{+0.2}_{-0.4}$ (all errors are 1$\sigma$) 
and for EMNO-7 we found a photometric redshift of $z=2.5^{+0.2}_{-0.8}$.
We also searched for secondary redshift solutions and for EMNO-3 and EMNO-7 we 
found secondary solutions that were within the 1$\sigma$ error bars of the 
primary solutions.
For EMNO-5 the photometric redshift was better constrained with a very well fitted 
Arp220-like SED at a redshift of $z=0.9^{+0.1}_{-0.2}$ all the way to radio
wavelengths (this is one of only two of our EMNOs also having a radio
detection) making this source a good candidate for a high-redshift Arp220 ULIRG
clone. For EMNO-5 the secondary redshift solution $z_{sec}=1.4$ had a 
confidence level that was lower by a factor of ten than the confidence level of the 
primary solution.

For the four objects with no optical detections we selected the most
'conservative' redshift and SED combination (i.e.\ starting from
low-redshift M82 SEDs) fitting the optical and FIR upper limits when normalized
in the $K$-band.  
Three HR10 SEDs and one M82-type SED were fitted, i.e.\ all highly reddened starbursts at $z\sim
1.5$ (Table~\ref{table-phot}). With our current data we can conservatively
estimate our photometric redshifts to be accurate within $z=\pm0.5$ with a higher
confidence given to the objects with HYPERZ fits than the ones with only SED fits.

Interestingly, however, the FIR (and radio) upper limits of most red
EMNOs rule out an Arp220 SED, which otherwise would explain most or
all of the observed MIR radiation.  This is very similar to the recent
Haas et al. (\cite{Ha04}) findings where sources selected by high
$f_{6.7} / f_{K}$ ratio have low $f_{60} / f_{MIR}$, indicative of 
AGN.  Our SEDs are clearly suggestive of hotter dust as well, as seen
from the overplotted appropriately
redshifted and arbitrarily normalized nuclear 
SED of the well-known Sy-2 NGC1068 (Fig.~\ref{emno_seds}, thick
line).  EMNO-6 has 
additionally been overplotted with an example of Treister et al.\
(\cite{Tr04}) 
obscured AGN model with $N_H = 10^{23} \ {\rm cm}^{-2}$ and $L_x =
  10^{43.5} \ {\rm  erg \  s^{-1}}$, which fits the red EMNOs, with
  the exception of EMNO-5, remarkably well.  In general,
the higher the MIR/NIR ratio is, the more likely and dominating the
AGN component will be, as already suggested above in
Section~\ref{definition}.  For 
example, in the sample of Prouton et al. (\cite{Pr04}), the only ULIRG
that would be classified as an EMNO is their only one where close to
$\sim50$\% of the IR luminosity is provided by the AGN component.  

We do note that extreme starformation alone in obscured
{\em very young} starburst regions can also produce steep MIR SEDs
without yet contradicting FIR upper limits (Takeuchi et
al. \cite{Ta03}).  At $z>1$ the red EMNOs would truly need to be
massive primeval galaxies 
undergoing their very first starburst, and we find the AGN
contribution option to be the more conservative one.  

The blue EMNOs have optical colours of a less
extincted younger stellar population.  The overplotted SEDs are those
of starforming GRASIL spirals (Table~\ref{table-phot}), which
appear to need large contributions 
to MIR flux from obscured central engines.  Accordingly, EMNO-9 is
a spectroscopically confirmed ($z=2.610$) moderately reddened QSO with 
very large X-ray absorbtion ($N_H \approx 3\times10^{23} \ {\rm
  cm}^{-2}$; 
Willott et al.\ \cite{Wi03}; object N2\_25 therein). All of our other
EMNOs unfortunately fall out of 
the X-ray coverage of ELAIS, but the optical to MIR
characteristics of EMNO-7, with a photometric redshift of $z=2.5^{+0.2}_{-0.8}$
(HYPERZ $1\sigma$ errors, $K$-band not used since
there is likely AGN contribution), look very much like that of
EMNO-9.  

EMNO-8 is a peculiar case, with a spectroscopic redshift of only
$z=0.1882$, making its luminosity lower than the other EMNOs by
two orders of magnitude.  Optical and NIR sizes indicate a dwarf,
$r_{hl} \approx 2$ kpc.  Interestingly, its SED follows well that of 
SBS 0335-052 (Dale et al. \cite{Da01}) 
an extremely metal poor dwarf galaxy, whose
surprisingly strong mid-IR flux is interpreted to result from hot dust
heated in an extremely intense interstellar radiation field around a
few Myr old starforming region.  Our EMNO-8 is approximately 10 times
more luminous, and whether its properties are a scaled up version of
SBS 0335-052 or whether there is some AGN activity involved, remains
to be studied.  Puzzlingly, the ELAIS Band-Merged Catalogue 
(Rowan-Robinson et al. \cite{Row04}) does not mention any
signs of starburst or AGN signatures in its optical spectrum.
Finally, we note that an $r-K>6.5$ ERO is just within the $3\sigma$
error circle of the ISOCAM position -- if it were the true
counterpart, the SED and interpretation would be identical to the
red EMNOs.

\subsection{Star-formation rates}

Due to their extreme nature, star-formation rates (SFR) based on
indicators used in the literature cannot be confidently applied to our
EMNOs.  The following should therefore be regarded as an attempt to
obtain indicative 
levels of SFR for the EMNOs.  First, total IR luminosity is
calculated from the best fit GRASIL SEDs of each object and by scaling
the SED to observed $K$-band luminosity.  Following Mann et
al. (\cite{Ma02}) we then derive  ${\rm SFR \ [M_{\sun} / yr]} = 
  L_{\rm IR} {\rm (3-1000\mu { m}) / [2.3  \times 10^{36}  
  W]^{1.05} } $. These are the $L_{IR}$ and SFR values listed in
Table~\ref{table-phot}.  SFRs using SEDs of M82 and Arp220 bracket
values of $\sim 200 - 1000 \ {\rm M_{\sun}/ yr}$. 

Similarly, normal late type spiral SEDs approximately fitting the blue
EMNOs result in SFRs of the order of $\sim 100 {\rm M_{\sun}/ yr}$ for
the high redshift ones, and $<1 \ {\rm M_{\sun}/ yr}$ for EMNO-8.

Had we scaled the model SEDs to the detected ISOCAM fluxes, the
SFRs would have been higher by factors of 2--5.
Similarly, if we directly used the 15$\mu$m flux as a SFR indicator
as calibrated by Elbaz et al. (\cite{El02a}, see their Fig.~5), our
red EMNOs would have very large SFR values at a level of 5000 ${\rm
  M_{\sun}/ yr}$, and the high-$z$ blue EMNOs 3--5 times higher than
these.  This estimate is of course incorrect if, as
suggested by the SED shapes, most of the 15 $\mu$m flux density comes
from AGN related hot dust emission rather than from star formation.
However, the more conservative values obtained above might not be any
more accurate, EMNOs being very extreme objects with regard to their
SED shape in the near to mid-IR. 

Finally, it is important to note that also fully evolved ellipticals
at $z\sim$1--2 satisfy very well our optical-NIR colour limits. In
this  
case {\em all} the mid-IR flux would have to be obscured AGN
originated -- to illustrate this, the SEDs of EMNOs 1 and 6 are
overplotted (dashed lines) with a GRASIL elliptical, and
AGN+elliptical model from Treister et al. (\cite{Tr04}), respectively.
Nearly all red EMNOs are consistent with
this old elliptical host + AGN possibility, and thus the SFR {\em lower
  limit} for 5 out of 6 of these is essentially zero.
From the elliptical GRASIL model we estimate that if the $K$-band
light of the red EMNOs is assumed to be totally of stellar origin,
their stellar masses are of the order of 0.5-1$\cdot 10^{12} \ {\rm
  M_{\sun}}$, implying black hole masses of $>10^9 \ {\rm M_{\sun}}$  
(Marconi \& Hunt \cite{Ma03}).  

\subsection{Number counts and contribution to the extragalactic
  background} 

Proper number densities cannot be derived because of a targeted and
discontinuous survey area.  However, we estimate a definite lower
limit on the number of mid-IR selected EMNOs in the following way.
In our IRTF follow-up 12 ISO sources (with $\approx$1--3 mJy) were
observed, 6 of which were 
EMNOs (and 3 of which red EMNOs).  The ISO sources were selected
because of their faint or non-existent counterparts in DSS, thus
obviously biasing us to large MIR to optical-NIR flux ratios.
These ISOCAM sources were spread
around an area of 0.15 deg$^2$, which {\em altogether} contains
30 $15\mu$m sources in the range 1--3 mJy, including the
ones observed.  Thus, assuming that {\em none} of the non-observed MIR
sources are EMNOs, EMNOs make up 20\% of all $15\mu$m sources at
$\sim1$ mJy.  If the targeted 12 sources are representative, on the
other hand, the EMNO fraction could be even 50\%.  In contrast, less
than 1\% of the general ELAIS catalogue sources (all fluxes) are
EMNOs (Fig.~\ref{iso_comb}).  In the selected deeper (in MIR) ISO
surveys (Fig.~\ref{iso_comb}), $<10$\% of the sources are EMNOs -- 
however, these surveys have widely varying depths of NIR coverage.

In terms of total flux density, all the ISOCAM sources brighter than 1
mJy (excluding a bright star) in the IRTF survey area amount to
$\approx 0.3 \ {\rm nW \ m^{-2} \ sr^{-1}}$, i.e.\ 10\% of the
total 
CIRB at 15 $\mu$m, according to Elbaz et al.\ (\cite{El02a}), totally
consistent with values therein.  EMNOs contribute 22\% of the flux of
all the ISOCAM sources in the area, which thus is the minimum EMNO
contribution to CIRB at this flux level.  
As above, if the 6 EMNOs on the other hand
were representative of the area, up to half of the CIRB is
contributed by EMNOs.   Fadda et al.\ (\cite{Fa02}) find a
$17\pm7$\% contribution by AGN to MIR background at these flux
levels.  As we have argued, EMNOs are likely to be dominated by AGN
activity.  Assuming full EMNO/AGN correspondence, our value for AGN
contribution is thus consistent with theirs. If the fraction of CIRB
contribution using our method turns out to be larger, up to 50\% as is
possible, this could suggest that X-ray surveys are indeed missing the
most obscured AGN, and MIR surveys could help in this regard (see
e.g.\ Haas et al. \cite{Ha04}; Gandhi \& Fabian \cite{Ga03});
Compton thick AGN would be obscured even in hard X-rays while their
nuclear SED in NIR-MIR should still have EMNO characteristics.
At the moment we cannot check this speculation because of lacking
X-ray observations of the sources.

\section{Conclusions}

We have presented a population of sources characterized by very strong
MIR flux compared to their optical and NIR properties.  
The sources have been dubbed EMNOs, and are argued to host an obscured AGN
based on: i) Their spectral shape in the near to far-IR region which
shows a high MIR/NIR flux ratio and in most cases a low FIR/MIR
ratio.  ii) The high-redshift spectroscopically confirmed EMNO is
shown to be a heavily obscured quasar by X-ray data (others do not have
X-ray coverage). iii) All the objects found from several other ISO
surveys which satisfy our EMNO criterion are very hard X-ray sources
suggesting obscured AGN.

Those of our EMNOs which are red in their optical-NIR colours are well 
understood as $z=1-1.5$ (ultra)luminous dusty galaxies based on
their SEDs and photometric redshifts.  Their mid-IR flux density is
not explained, however, without a dominant fraction of AGN
activity.  Moreover, at least for 4 out of 6 of these cases are also
consistent with passive early type galaxy SEDs, in which case {\em
  all} the mid and far-IR radiation would be AGN produced.

The blue EMNOs appear to be higher redshift obscured quasars, according
to the one spectroscopically confirmed case.  However, their optical
to FIR broadband colours appear to be very similar to low redshift
extreme starbursting dwarfs, based on the similarity of our EMNO-8 and
the well-studied metal-poor dwarf SBS 0335-052.  Spectra, or x-ray and 
radio data are needed to study this degeneracy further.  

EMNOs contribute 20--50\% of the extragalactic source counts and
cosmic IR background radiation at the 1 mJy level.  If further
observations show them to be dominated by AGN, as we suspect, it could
mean that a significant fraction of AGN activity is missed by X-ray
surveys.

We predict that the EMNO criterion will be
very useful in selecting obscured AGN and sources with extremely
strong starformation, but also in disentangling the AGN from
starbursts since the MIR/NIR flux ratio is larger for the AGN with
appropriately selected bands.  Spitzer surveys using IRAC and
MIPS bands (such as SWIRE; Lonsdale et al. \cite{Lo03}) are expected
to detect large numbers of EMNOs.  The inclusion of MIPS 70$\mu$m data
will be particularly useful in constraining the AGN/starburst fraction
in EMNOs and generally in better classifying this new population.

\begin{acknowledgements}

We thankfully acknowledge the work done by the ELAIS collaboration
and ISO science centre in bringing together the ELAIS products. 
We wish to warmly thank Alberto Franceschini, Gian Luigi Granato,
Laura Silva, Poshak Gandhi, and Valentin D.\ Ivanov for good
suggestions and discussion, as well as the referee for very
constructive remarks.  We also thank Granato and Silva, and
Ezekiel Treister for providing their models.

\end{acknowledgements}

\end{document}